\documentclass[twocolumn,showpacs,preprintnumbers,superscriptaddress,amsmath,amssymb]{revtex4}

\usepackage{graphicx}
\usepackage{dcolumn}
\usepackage{bm}

\begin{document}

\title{Evolving Apollonian networks with small-world scale-free topologies}

\author{Zhongzhi Zhang}
\affiliation{Department of Computer Science and Engineering, Fudan
University, Shanghai 200433, China}%
\email{xinjizzz@sina.com (Z.Z. Zhang), sgzhou@fudan.edu.cn (S. G.
Zhou)}

\author{Lili Rong}
\affiliation{Institute of Systems Engineering, Dalian University of
Technology, Dalian 116023, Liaoning, China}%
\email{llrong@dlut.edu.cn}

\author{Shuigeng Zhou}
\affiliation{Department of Computer Science and Engineering, Fudan
University, Shanghai 200433, China}%
\homepage{http://www.iipl.fudan.edu.cn/~zhousg/sgzhou.htm}
\email{sgzhou@fudan.edu.cn}

\date{\today}

\begin{abstract}
We propose two types of evolving networks: evolutionary Apollonian
networks (EAN) and general deterministic Apollonian networks (GDAN),
established by simple iteration algorithms. We investigate the two
networks by both simulation and theoretical prediction. Analytical
results show that both networks follow power-law degree
distributions, with distribution exponents continuously tuned from 2
to 3. The accurate expression of clustering coefficient is also
given for both networks. Moreover, the investigation of the average
path length of EAN and the diameter of GDAN reveals that these two
types of networks possess small-world feature. In addition, we study
the collective synchronization behavior on some limitations of the
EAN.
\end{abstract}


\maketitle


\section{Introduction}

In the last few years, complex networks have attracted a growing
interest from a wide circle of researchers
\cite{Ba02,DoMe03,SaVe04,Ne03}, with particular focus on the
following three properties: power-law degree distribution
\cite{BaAl99}, small average path length (APL) and high clustering
coefficient \cite{WaSt98}. The reason for this boom is that complex
networks describe many systems in nature and society which sharing
the above-mentioned three characteristics.

In order to mimic the real-world systems, a wide variety of models
have been proposed~\cite{Ba02,DoMe03,SaVe04,Ne03}. The first
successful attempt is the Watts and Strogatz's small-world network
model (WS model) \cite{WaSt98}, which started an avalanche of
research on the models of small-world networks and the WS model
\cite{NeWa99a, DoMe00, CoOzPe00, OzHuOt04, BlKr05}. Another
well-known model is Barab\'asi and Albert's elegant scale-free
network model (BA model) ~\cite{BaAl99}, which has attracted an
exceptional amount of attention within the physics community. In
addition to analytic studies of the BA model and research of its
extensions or modifications~\cite{DoMeSa00,KaReLe00,Frfrho03}, many
authors have developed a considerable number of other models and
mechanisms that may represent processes more realistically taking
place in real-world
networks~\cite{AmScBaSt00,BiBa01,AlBa00,DoMe00b,ChLuDeGa03,Ke04,WWX1,WWX2,WWX3}.
Until now, modeling complex networks with small-world and scale-free
properties is still an important issue.

Recently, based on the well-known Apollonian packing, Andrade
\emph{et al.} introduced Apollonian networks~\cite{AnHeAnSi05} which
were also proposed by Doye and Massen in~\cite{DoMa05}
simultaneously. Apollonian networks belong to a deterministic
growing type of networks, which have drawn much attention from
scientific
communities~\cite{BaRaVi01,IgYa05,DoGoMe02,JuKiKa02,CoFeRa04,RaBa03,No03,ZhWaHuCh04,ZhRoGo05,ZhRoCo05a}.
The effects of the Apollonian networks on several dynamical models
have been intensively studied, including Ising model and a magnetic
model \emph{et al}~\cite{AnHeAnSi05,AnHe05,LiGaHe04,AnMi05}. Doye
and Massen adopted an extension~\cite{DoMa05} of two-dimensional to
investigate energy landscape networks~\cite{Do02,DoMa05b}. Zhang
\emph{et al} proposed a minimal iterative algorithm for constructing
high dimensional networks and studied their structural properties
\cite{ZhCoFeRo05}. In \cite{ZhYaWa05} Zhou \emph{et al} proposed a
simple rule that generates random two-dimensional Apollonian
networks, which are generalized by Zhang \emph{et al} to high
dimension \cite{ZhRoCo05}. The deterministic Apollonian networks
(DAN)~\cite{AnHeAnSi05,DoMa05,ZhCoFeRo05} and random Apollonian
networks (RAN) \cite{ZhYaWa05,ZhRoCo05} may provide valuable
insights into the real-life networks.

In this paper, first we propose a general scenario for constructing
evolutionary Apollonian networks (EAN) controlled by a parameter
$q$. The EAN can also result from Apollonian packing and unifies the
DAN and RAN to the same framework, i.e., the DAN and RAN are special
cases of EAN. Then, we present a general deterministic Apollonian
network (GDAN) model governed by a single parameter $m$. Both EAN
and GDAN have a power-law degree distribution, a very large
clustering coefficient and a small intervertex separation. The
degree exponent of EAN and GDAN is changeable between 2 and 3.
Moreover, we introduce an algorithm for the DAN and RAN which can
realize the construction of our networks. In the end, through
eigenvalue spectrum of the Laplacian matrix, the synchronizability
on some limiting cases of the EAN is discussed.

\section{Brief introduction to deterministic and random Apollonian networks}
%
We first introduce Apollonian packing (see Fig. 1 for an example of
two dimension), which dates back to ancient Greek mathematician
Apollonius of Perga. The classic two-dimensional Apollonian packing
is constructed as follows. Initially three mutually touching disks
are inscribed inside a circular space which is to be filled. The
interstices of the initial disks and circle are curvilinear
triangles to be filled. We denote this initial configuration by
generation $t=0$. Then in the first generation $t=1$, four disks are
inscribed, each touching all the sides of the corresponding
curvilinear triangle. The process is repeated indefinitely for all
the new curvilinear triangles. In the limit of infinite generations,
we obtain the well-known two-dimensional Apollonian packing. The
translation from Apollonian packing construction to Apollonian
network generation is quite straightforward. Let the nodes
(vertices) of the network correspond to the disks and make two nodes
connected if the corresponding disks are
tangent~\cite{AnHeAnSi05,DoMa05}. Fig. 2 shows a network based on
the two-dimensional Apollonian packing.
\begin{figure}
\begin{center}
\includegraphics[width=0.6\textwidth]{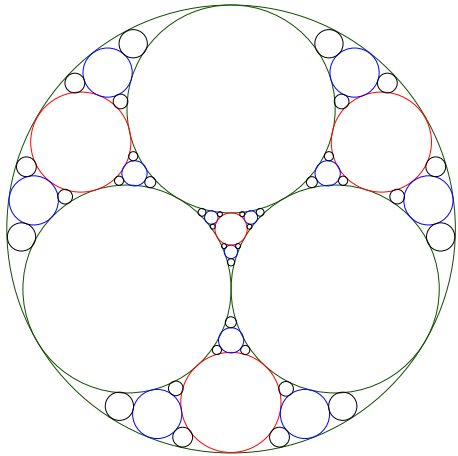} \
\end{center}
\caption[kurzform]{\label{Fig1} (Color online) A two-dimensional
Apollonian packing of disks. }
\end{figure}
The two-dimensional Apollonian network can be easily generalized to
high-dimensions ($d$-dimensional, $d\geq 2$)
\cite{DoMa05,ZhCoFeRo05} associated with other self-similar packings
\cite{MaHeRi04}. The $d$-dimensional Apollonian packings start with
$d+1$ mutually touching $d$-dimensional hyperspheres that is
enclosed within and touching a larger $d$-dimensional hyperspheres,
with $d+2$ curvilinear $d$-dimensional simplex ($d$-simplex) as
their interstices, which are to be filled in successive generations.
If each $d$-hypersphere corresponds to a node and nodes are
connected if the corresponding $d$-hyperspheres are in contact, then
$d$-dimensional Apollonian networks are established. In every
generation of the $d$-dimensional Apollonian packings, if we add
only one $d$-hypersphere inside a randomly selected interstice, then
we get a high dimensional random Apollonian packings, based on which
high dimensional random Apollonian networks are constructed
\cite{ZhYaWa05,ZhRoCo05}.
\begin{figure}
\begin{center}
\includegraphics[width=0.4\textwidth]{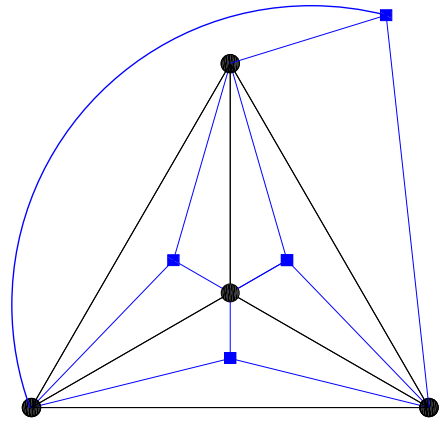} \
\end{center}
\caption[kurzform]{\label{network} (Color online) Illustration of
the two-dimensional deterministic Apollonian networks, showing the
first two steps of iterative process. }
\end{figure}

\section{The iterative algorithms for the networks}
%

Before introducing the algorithms we give the following definitions
on a graph. The term \emph{size} refers to the number of edges in a
graph. The number of nodes in the graph is called its \emph{order}.
When two nodes of a graph are connected by an edge, these nodes are
said to be \emph{adjacent}, and the edge is said to join them. A
\emph{complete graph} is a graph in which all nodes are adjacent to
one another. Thus, in a complete graph, every possible edge is
present. The complete graph with $d$ graph nodes is denoted as $K_d$
(also referred in the literature as $d$-\emph{clique}; see
\cite{We01}). Two graphs are \emph{isomorphic} when the nodes of one
can be relabeled to match the nodes of the other in a way that
preserves adjacency. Hence all $d$-cliques are isomorphic to one
another.

\subsection{Evolutionary Apollonian networks}
Now we introduce the evolving Apollonian networks. First we give a
new packing method for high-dimensional ($d$-dimensional, $d\geq 2$)
Apollonian packings. The initial configuration is the same as the
deterministic Apollonian packings. Then in each subsequent
generation, each $d$-simplex is filled with probability $q$. In a
special case $q=1$, it is reduced to the classic deterministic
Apollonian packings. if $q$ approaches but is not equal to 0, it
coincides with the random Apollonian packings described in
\cite{ZhYaWa05,ZhRoCo05}. The EAN is derived from this new packing:
nodes represent $d$-hyperspheres and edges correspond to contact
relationship. Fig. 2 shows the network growing process for a special
case of $d=2$ and $q=1$.

In the construction process of the new high-dimensional Apollonian
packings, for each new $d$-hypersphere, $d+1$ new interstices are
created. When building networks, it is equivalent that for each new
added node, $d+1$ new $d$-simplices are generated, which may create
new nodes in the subsequent generations. According to this, we can
introduce a general algorithm to create EAN, using which one can
write computer program conveniently to simulate the networks and
study their properties.

The $d$-dimensional EAN after $t$ generations are denoted by
$A(d,t)$, $d\geq 2, t\geq 0$. Then at step $t$, the $d$-dimensional
EAN is constructed as follows: For $t=0$, $A(d,0)$ is a complete
graph $K_{d+2}$ (or $(d+2)$-clique). For $t\geq 1$, $A(d,t)$ is
obtained from $A(d,t-1)$. For each of the existing subgraphs of
$A(d,t-1)$ that is isomorphic to a $(d+1)$-clique and has {\em never
generated a node before} (we call them active $(d+1)$-cliques), with
probability $q$, a new node is created and connected to all the
nodes of this subgraph. The growing process is repeated until the
network reaches a desired order. When $q=1$, the networks are
exactly the same as the DAN~\cite{AnHeAnSi05,DoMa05, ZhCoFeRo05}. If
$q<1$, the networks are growing randomly. Especially, as $q$
approaches zero and does not equal to zero, the networks are reduced
to the RAN studied in detail in~\cite{ZhYaWa05,ZhRoCo05}.
See~\cite{Do03} for interpretation.

Next we compute the size and order of EAN. Note that the addition of
each new node leads to $d+1$ new $(d+1)$-cliques and $d+1$ new
edges. Then, at step 1, we add expected $L_v(1)=(d+2)q$ new nodes
and $L_e(1)=(d+2)(d+1)q$ new edges to the graph. After simple
calculations, one can obtain that at $t_i$($t_i>1$) the numbers of
newly born nodes and edges are $L_v(t_i)=q(d+2)(1+dq)^{t_i-1}$ and
$L_e(t_i)=q(d+1)(d+2)(1+dq)^{t_i-1}$, respectively. Thus the average
number of total nodes $N_t$ and edges $E_t$ present at step $t$ is
\begin{eqnarray}\label{Nt1}
N_t&=&(d+2)+\sum_{t_i=1}^{t}L_v(t_i)\nonumber\\
&=&(d+2)\frac {(1+dq
)^{t}+d-1}{d}
\end{eqnarray}
and
\begin{eqnarray}\label{Et1}
E_t &=&\frac{(d+2)(d+1)}{2}+
\sum_{t_i=1}^{t}L_e(t_i)\nonumber\\
&=&(d+2)(d+1) \frac{2(1+dq )^{t}+d-2}{2d},
\end{eqnarray}
respectively. So for large $t$, The average degree $\overline{k}_t=
\frac{2E_t}{N_t}$ is approximately $2(d+1)$. Moreover, when $d=2$,
we have $E_t=3N_{t}-6$. Thus in this case, all networks are maximal
planar networks (or graphs)~\cite{ZhYaWa05}.

\subsection{General deterministic Apollonian networks}
According to the construction process of $d$-dimensional Apollonian
packings, in Ref.~\cite{ZhCoFeRo05}, a generation algorithm for
$d$-dimensional Apollonian networks was proposed. Here we generalize
the algorithm to establish general deterministic Apollonian networks
(GDAN). The network, denoted by $G(d,t)$ after $t$ generations with
$d\geq 2$ and $t\geq 0$, is constructed in an iterative way. For
$t=0$, $G(d,0)$ is a complete graph $K_{d+2}$ (or $(d+2)$-clique).
For $t\geq 1$, $G(d,t)$ is obtained from $G(d,t-1)$. For each of the
existing subgraphs of $G(d,t-1)$ that is isomorphic to a
$(d+1)$-clique and created at step $t-1$, $m$ new vertices are
created, and each is connected to all the vertices of this subgraph.
In the limit of infinite generations we obtain GDAN. When $m=1$, the
networks are reduced to DAN~\cite{AnHeAnSi05,DoMa05,ZhCoFeRo05}.

Let $L_v(t)$, $L_e(t)$ and $K_{(d+1),t}$ be the numbers of vertices,
edges and $(d+1)$-cliques created at step $t$, respectively. Because
the addition of each new vertex leads to $d+1$ new $(d+1)$-cliques
and $d+1$ new edges, we have $L_v(t)=mK_{(d+1),t-1}$,
$L_e(t)=(d+1)L_v(t)$,  and $K_{(d+1),t}=(d+1)L_v(t)$. Thus one can
easily obtain $K_{(d+1),t}=m(d+1)K_{(d+1),t-1}=(d+2)[m(d+1)]^{t}$
($t\geq0$), $L_v(t)=(d+2)m^{t}(d+1)^{t-1}$ ($t>0$) and
$L_e(t)=(d+2)m^{t}(d+1)^{t}$ ($t>0$). From these results, we can
easily compute the size and order of the networks. The total number
of vertices $N_t$ and edges $E_t$ present at step $t$ is
\begin{eqnarray}\label{Nt2}
N_t&=&\sum_{t_i=0}^{t}n_v(t_i)\nonumber\\
&=&\frac{m(d+2)[m^t(d+1)^{t}-1]}{m(d+1)-1}+d+2
\end{eqnarray}
and
\begin{eqnarray}\label{Et2}
E_t
=\sum_{t_i=0}^{t}n_e(t_i)\nonumber\\
=\frac{m(d+2)(d+1)[m^t(d+1)^{t}-1]}{m(d+1)-1}+\frac{(d+2)(d+1)}{2},
\end{eqnarray}
respectively. So for large large, The average degree
$\overline{k}_t= \frac{2E_t}{N_t}$ approaches $2(d+1)$.

\section{Relevant characteristics of the networks}
In the following we will study the topology properties of EAN and
GDAN, in terms of the degree distribution, clustering coefficient,
average path length and diameter.

\subsection{Degree distribution}
The degree distribution is one of the most important statistical
characteristics of a network. We will discuss the degree
distribution of EAN and GDAN in detail.

\subsubsection{\textbf{Evolutionary Apollonian networks}}
\begin{figure}
\begin{center}
\includegraphics[width=0.48\textwidth]{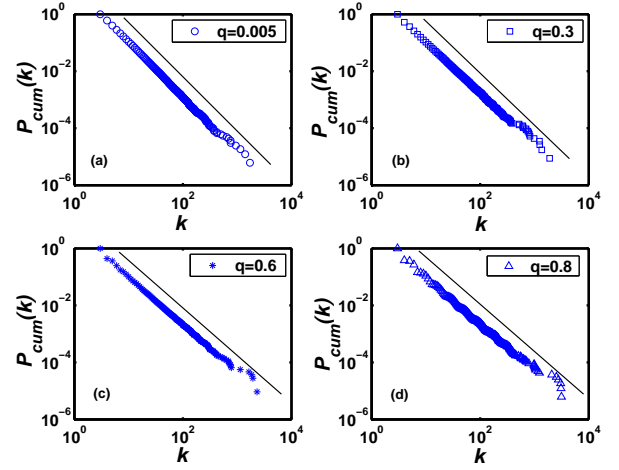} \
\end{center}
\caption[kurzform]{\label{Fig3} (Color online) The cumulative degree
distribution $P_{cum}(k)$ at various $q$ values for dimension $d=2$.
The circles (a), squares (b), stars (c), and triangles (d) denote
the simulation results for networks with different evolutionary
steps $t=1110$, $t=23$, $t=14$, and $t=12$, respectively. The four
straight lines are the theoretical results of $\gamma(d,q)$ as
provided by equation (6). All data are from the average of ten
independent simulations.}
\end{figure}

When a new node $i$ is added to the graph at step $t_i$, it has
degree $d+1$ and forms $d+1$ active $(d+1)$-cliques. Let $L_c(i,t)$
be the number of active $(d+1)$-cliques at step $t$ that will
possibly created new nodes connected to the node $i$ at step $t+1$.
Then at step $t_i$, $L_c(i, t_i)=d+1$. At step $t_i+1$, there are
$(d+1)q$ new nodes which forms $(d+1)qd$ new active $(d+1)$-cliques
containing $i$, and there are $(d+1)q$ active $(d+1)$-cliques of $i$
are deactivated at the same time. Thus $L_c(i,
t_i+1)=(d+1)[1+(d-1)q]$. It is not difficult to find the following
relation: $L_c(i, t+1)= L_c(i, t)[1+(d-1)q]$. Since $L_c(i,
t_i)=d+1$, we have $L_c(i, t)=(d+1)[1+(d-1)q]^{t-t_i}$. Then the
degree $ k_i(t)$ of node $i$ at time $t$ is
\begin{eqnarray} \label{Ki1}
k_i(t)&=&d+1+q\sum_{\tau=t_i}^{t-1}{ L_c(i,\tau)}\nonumber\\
&=&(d+1)\left(\frac{[1+(d-1)q]^{t-t_i}+d-2}{d-1}\right).
\end{eqnarray}
Since the degree of each node has been obtained explicitly as in
Eq.~(\ref{Ki1}), we can get the degree distribution via its
cumulative distribution, i.e., $P_{cum}(k) \equiv \sum_{k^\prime
\geq k} N(k^\prime,t)/N_t \sim k^{1-\gamma}$, where $N(k^\prime,t)$
denotes the number of nodes with degree $k^\prime$. The detailed
analysis is given as follows. For a degree $k$
\begin{equation*}
k=(d+1)\left(\frac{[1+(d-1)q]^{t-m}+d-2}{d-1}\right),
\end{equation*}
there are  $L_v(m)=q(d+2)(1+dq)^{m-1}$ nodes with this exact degree,
all of which were born at step $m$. All nodes born at time $m$ or
earlier have this or a higher degree. So we have
\begin{eqnarray}
\sum_{k' \geq k}
N(k',t)&=&(d+2)+\sum_{s=1}^{m}L_v(s)\nonumber\\
&=&(d+2)\frac {(1+dq )^{m}+d-1}{d}\nonumber.
\end{eqnarray}
As the total number of nodes at step $t$ is given in Eq.~(\ref{Nt1})
we have
\begin{eqnarray}
\left [(d+1)\left(\frac{[1+(d-1)q]^{t-m}+d-2}{d-1}\right)\right]^{1-\gamma}\nonumber\\
=\frac{(d+2)\frac {(1+dq )^{m}+d-1}{d}}{(d+2)\frac {(1+dq
)^{t}+d-1}{d}}\nonumber.
\end{eqnarray}
Therefore, for large $t$ we obtain
\begin{equation*}
\left[[1+(d-1)q]^{t-m}\right]^{1-\gamma}=(1+dq)^{m-t}
\end{equation*}
and
\begin{equation}
\gamma \approx 1+\frac{\ln (1+dq)}{\ln[1+(d-1)q]}.
\end{equation}
Thus, the degree exponent $\gamma$ is a continuous function of $d$
and $q$, and belongs to the interval (2,3]. For any fixed $d$, as
$q$ decrease from 1 to 0, $\gamma$ increases from $1+\frac{\ln
(1+d)}{\ln d}$ to $2+\frac{1}{d-1}$. In the case $d=2$, $\gamma$ can
be tunable between 2.58496 and 3. In the two limitations, i.e.,
$q=1$ and $q\rightarrow0$ (but $q \neq0$), the evolutionary
Apollonian network reduces to the deterministic Apollonian
networks~\cite{AnHeAnSi05,DoMa05,ZhCoFeRo05} and their stochastic
variants \cite{ZhYaWa05,ZhRoCo05}, respectively. Fig.~(\ref{Fig3})
shows, on a logarithmic scale, the scaling behavior of the
cumulative degree distribution $P_{cum}(k)$ for different values of
$q$ in the case of $d=2$. Simulation results agree very well with
the analytical ones.

\subsubsection{\textbf{General deterministic Apollonian networks}}
Considering a vertex $i$ added to the networks at step $t_i$. Let
$K_{(d+1)}(i,t)$ be the number of newly-created $(d+1)$-cliques at
step $t$ containing vertex $i$. These new cliques will create new
vertices connected to the vertex $i$ at step $t+1$. At step $t_i$,
$K_{(d+1)}(i,t_i)=d+1$. From the iterative process, one can see that
each new neighbor of $i$ generates $d$ new $(d+1)$-cliques with $i$
belonging to them. Then it is not difficult to find following
relations:
\begin{equation}
\Delta k_i(t)=k_i(t)-k_i(t-1)=mK_{(d+1)}(i,t-1)\nonumber\\
\end{equation}
and
\begin{equation}
K_{(d+1)}(i,t)=dmK_{(d+1)}(i,t-1)=(d+1)(dm)^{t-t_{i}}\nonumber\\.
\end{equation}
Then the degree of vertex $i$ becomes
\begin{eqnarray}\label{Ki2}
k_i(t)&=&k_i(t_i)+m\sum_{\tau=t_i}^{t-1}
K_{(d+1)}(i,\tau)\nonumber\\&=&\frac{m(d+1)[(md)^{t-t_{i}}-1]+d^{2}-1}{d-1}.
\end{eqnarray}

For a degree $k$
\begin{equation*}
k=\frac{m(d+1)[(md)^{t-p}-1]+d^{2}-1}{d-1},
\end{equation*}
there are  $n_v(p)=(d+2)m^{p}(d+1)^{p-1}$ vertices with this exact
degree. Also, we have
\begin{eqnarray}
\sum_{k' \geq k}
N(k',t)=\sum_{s=0}^{p}n_v(s)\nonumber\\
=\frac{m(d+2)[m^{p}(d+1)^{p}-1]}{m(d+1)-1}+d+2.
\end{eqnarray}
From the definition of cumulative degree distribution, we have
\begin{eqnarray}
\left[\frac{m(d+1)[(md)^{t-p}-1]+d^{2}-1}{d-1}\right]^{1-\gamma}\nonumber\\
=\frac{\frac{m(d+2)[m^{p}(d+1)^{p}-1]}{m(d+1)-1}+d+2}{\frac{m(d+2)[m^t(d+1)^{t}-1]}{m(d+1)-1}+d+2}\nonumber.
\end{eqnarray}
When $t$ is large enough, one can obtain
\begin{equation*}
\left[(md)^{t-p}\right]^{1-\gamma}=[m(d+1)]^{p-t}
\end{equation*}
and
\begin{equation}\label{Gamma}
\gamma \approx 1+\frac{\ln [m(d+1)]}{\ln(md)}.
\end{equation}
For $m=1$, Eq. (\ref{Gamma}) recovers the results previously
obtained in Refs.~\cite{AnHeAnSi05,DoMa05,ZhCoFeRo05}.
\subsection{Clustering coefficient}

The clustering coefficient~\cite{WaSt98} $ C_i $ of node $i$ is
defined as the ratio between the number of edges $e_i $ that
actually exist among the $k_i $ neighbors of node $i$ and its
maximum possible value, $ k_i( k_i -1)/2 $, i.e., $ C_i =2e_i/[k_i(
k_i -1)]$. The clustering coefficient of the whole network is the
average of $C_i^{'}s $ over all nodes in the network.

For both EAN and GDAN, the analytical expression of clustering
coefficient $C(k)$ for a single node with degree $k$ can be derived
exactly. When a node is created it is connected to all the nodes of
a $(d+1)$-clique, in which nodes are completely interconnected. So
its degree and clustering coefficient are $d+1$ and 1, respectively.
In the following steps, if its degree increases one by a newly
created node connecting to it, then there must be $d$ existing
neighbors of it attaching to the new node at the same time. Thus for
a node of degree $k$, we have
\begin{equation}\label{Ck}
C(k)= {{{d(d+1)\over 2}+ d(k-d-1)} \over {k(k-1)\over 2}}=
\frac{2d(k-\frac{d+1}{2})}{k(k-1)},
\end{equation}
which depends on degree $k$ and dimension $d$. For $k \gg d$, the
$C(k)$ is inversely proportional to degree $k$. The scaling
$C(k)\sim k^{-1}$ has been found for some network
models~\cite{AnHeAnSi05,DoMa05,DoGoMe02,CoFeRa04,RaBa03,No03,ZhCoFeRo05,ZhYaWa05,ZhRoCo05},
including DAN and
RAN~\cite{AnHeAnSi05,DoMa05,ZhCoFeRo05,ZhYaWa05,ZhRoCo05}, and has
also been observed in several real-life networks~\cite{RaBa03}.

Using Eq. (\ref{Ck}), we can obtain the clustering $\overline{C}_t$
of the networks at step $t$:
\begin{equation}\label{ACCk}
\overline{C}_t=
    \frac{1}{N_{t}}\sum_{r=0}^{t} \frac{2d(D_r-\frac{d+1}{2})L_v(r)}{D_r(D_r-1)}
\end{equation}
where the sum runs over all the nodes
and $D_r$ 
is the degree of the nodes created at step $r$, which is given by
Eq. (\ref{Ki1}) or (\ref{Ki2}).

\subsubsection{\textbf{Evolutionary Apollonian networks}}
\begin{figure}
\begin{center}
\includegraphics[width=0.45\textwidth]{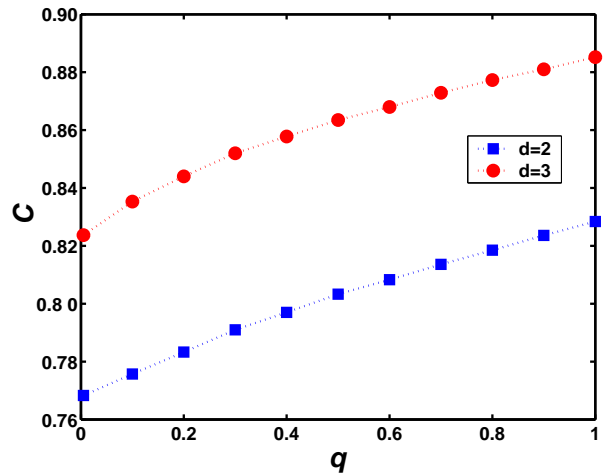} \
\end{center}
\caption[kurzform]{\label{ACC1} (Color online) The clustering
coefficient of the whole network as a function of $q$ and $d$.
Results are averaged over ten network realizations for each datum.}

\end{figure}

For EAN, it can be easily proved that for any fixed $d$,
$\overline{C}_t$ increases with $q$, and that for arbitrary fixed
$q$, $\overline{C}_t$ increases with $d$. Exactly analytical
computation shows: in the case $d=2$, when $q$ increases from 0 to
1, $\overline{C}$ grows from 0.7366 to 0.8284, with a special value
$\overline{C}_t=0.7934$ for $q=0.5$. Likewise, in the case $d=3$,
$\overline{C}$ increases from 0.8021 to 0.8852 when $q$ increases
from 0 to 1, especially $\overline{C}_t=0.8585$ for $q=0.5$ (see
also ~\cite{AnHeAnSi05,ZhCoFeRo05,ZhYaWa05,ZhRoCo05}). Therefore,
the evolutionary networks are highly clustered. Fig.~(\ref{ACC1})
shows the clustering coefficient of the network as a function of $q$
for $d=2$ and $d=3$, respectively, which is in accordance with our
above conclusions. From~(\ref{Fig3}) and Figs.~(\ref{ACC1}), one can
see that both degree exponent $\gamma$ and clustering coefficient
$\overline{C}_t$ depend on the parameter $q$. The mechanism
resulting in this relation deserves further study. The fact that a
biased choice of the cliques at each iteration may be a possible
explanation, see Ref.~\cite{CoRobA05}.
\subsubsection{\textbf{General deterministic Apollonian networks}}
\begin{figure}
\begin{center}
\includegraphics[width=0.45\textwidth]{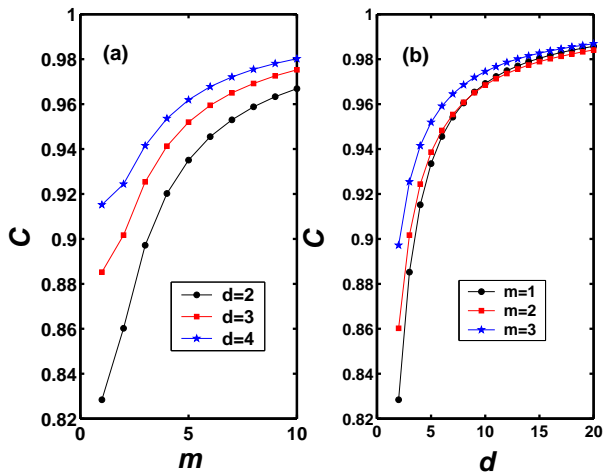} \
\end{center}
\caption[kurzform]{\label{ACC2} (Color online) The dependence
relation of $C$ on $d$ and $m$. }
\end{figure}

For GDAN, in the infinite network order limit ($N_{t}\rightarrow
\infty$), Eq. (\ref{ACCk}) converges to a nonzero value. When $d=2$,
for $m=1$, 2 and 3, $C$ is equal to 0.8284, 0.8602 and 0.8972,
respectively. When $m=2$, for $d=2$, 3 and 4, $C$ are 0.8602, 0.9017
and 0.9244, respectively. Therefore, the clustering coefficient of
GDAN is very high. Moreover, similarly to the degree exponent
$\gamma$, clustering coefficient $C$ is determined by both $d$ and
$m$. Fig. \ref{ACC2} shows the dependence of $C$ on $d$ and $m$.
From Fig. \ref{ACC2} (a) and (b), one can see that for any fixed
$m$, $C$ increases with $d$. But the dependence relation of $C$ on
$m$ (see Fig. \ref{ACC2} (b)) is more complex: (i) when $m\leq 2$
and $d\leq 6$, for the same $d$, $C$ increases with $m$; (ii) when
$m\leq 2$ and $d>6$, for the same $d$, $C$ decreases with $m$; (iii)
when $m\geq 3$, for arbitrary fixed $d$, $C$ increases with $m$.
Further effort should be paid for this complicated relation.

\subsection{Average path length for evolutionary Apollonian networks}
We label the nodes by their creation times, $v=1,2,3,\ldots,N-1,N.$
Using $\ell(N)$ to represent the APL of the our networks with order
$N$, then we have: $\ell(N)=\frac{2\sigma(N)}{N(N-1)}$, where
$\sigma(N)=\sum_{1 \leq i<j \leq N}d_{i,j}$ is the total distance,
in which $d_{i,j}$ is the smallest distance between node $i$ with
node $j$. Now we study the APL of the present model by using the
approach similar to that in ~\cite{ZhYaWa05,ZhRoCo05}.

Since the distances between existing node pairs will not be affected
by the addition of new nodes, thus we have:
\begin{equation}\label{E6}
\sigma(N+1) = \sigma(N)+ \sum_{i=1}^{N}d_{i,N+1}.
\end{equation}
Like in the analysis of ~\cite{ZhYaWa05,ZhRoCo05}, Eq.~(\ref{E6})
can be rewritten approximately as:
\begin{equation}\label{E7}
\sigma(N+1) = \sigma(N)+N+(N-d-1)\ell(N-d).
\end{equation}
It is clear
\begin{equation}\label{E8}
(N-d-1)\ell(N-d) = {2\sigma(N-d) \over N-d} < {2\sigma(N) \over N}.
\end{equation}
From Eqs.~(\ref{E7})and~(\ref{E8}), we can provide an upper bound
for the variation of $\sigma(N)$ as
\begin{equation}\label{E9}
{d\sigma(N) \over dN} =  N + {2\sigma(N) \over N},
\end{equation}
which leads to
\begin{equation}
\sigma(N) = N^2(\ln N + \beta),
\end{equation}
where $\beta$ is a constant. As $\sigma(N) \sim N^2\ln N $, we have
$\ell(N) \sim \ln N$.

\begin{figure}
\begin{center}
\includegraphics[width=0.45\textwidth]{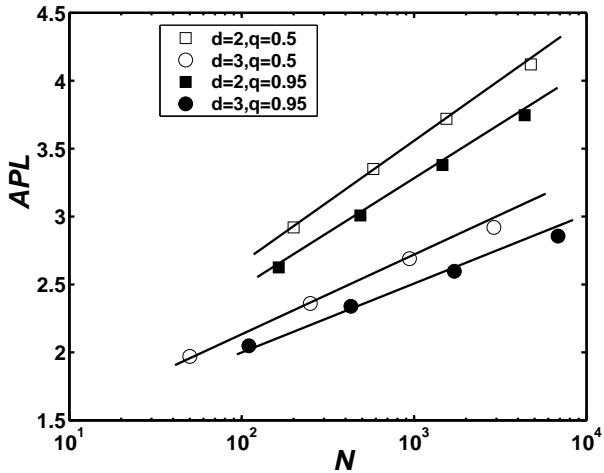} \
\end{center}
\caption[kurzform]{\label{Fig4} Semilogarithmic graph of the APL vs
the network order $N$. Besides $N$, APL depends on $d$ and $q$. Each
data is obtained by ten independent network realizations. The lines
are linear functions of ln N.}
\end{figure}

It should be emphasized that as Eq.~(\ref{E9}) has been deduced from
an inequality, then $\ell(N)$ increases at most as $\ln N$ with $N$.
Here we only give an upper bound for APL, which increases slightly
slower than $\ln N$. Thus, our model exhibits the small-world
property. Especially, in the case of $q=1$, we can compute exactly
the diameter of the network, which is the maximum distance between
all node pairs of a graph. Previously reported analytical result has
shown that the diameter grows logarithmically with the order of the
network~\cite{ZhCoFeRo05}. In Fig.~(\ref{Fig4}), we report the
simulation results on the APL of networks for different $q$ and $d$.
From Fig.~(\ref{Fig4}), one can see that for fixed $d$, APL
decreases with increasing $q$; and for fixed $q$, APL is a
decreasing function of $d$. When network order $N$ is small, APL is
a linear function of $lnN$; while $N$ becomes large, APL increases
slightly slower than $lnN$.

\subsection{Diameter for general deterministic Apollonian networks}
The diameter of a network characterizes the maximum communication
delay in the network and is defined as the longest shortest path
between all pairs of vertices. In what follows, the notations $
\lceil x \rceil$ and $\lfloor x \rfloor$ represent the integers
obtained by rounding $x$ to the nearest integers towards infinity
and minus infinity, respectively. Now we compute the diameter of
$A(d,t)$, denoted  by $diam(A(d,t))$ for $d\geq 2$:

{\em Step 0}. The diameter is $1$.

{\em Steps 1 to $\lceil\frac{d}{2}\rceil$}. In this case, the
diameter is 2, since any new vertex is by construction connected to
a $(d+1)$-clique, and since any $(d+1)$-clique during those steps
contains at least $\frac{d}{2}+2$ ($d$ even) or $\frac{d+1}{2}+1$
($d$ odd) vertices from the initial $(d+2)$-clique $A(d,0)$ obtained
after step 0. Hence, any two newly added vertices $u$ and $v$ will
be connected respectively to sets $S_u$ and $S_v$, with
$S_u\subseteq V(A(d,0))$ and $S_v\subseteq V(A(d,0))$, where
$V(A(d,0))$ is the vertex set of $A(d,0)$; however, since $\vert
S_u\vert\geq\frac{d}{2}+2$ ($d$ even) and $\vert
S_v\vert\geq\frac{d+1}{2}+1$ ($d$ odd), where $\vert S\vert$ denotes
the number of elements in set $S$, we conclude that $S_u\cap
S_v\neq{\O}$, and thus the diameter is 2.

{\em Steps
$\lceil\frac{d}{2}\rceil+1$ to $d+1$}. In any of those steps, some
newly added vertices might not share a neighbor in the original
$(d+2)$-clique $A(d,0)$ obtained after step 0; however, any newly
added vertex is connected to at least one vertex of the initial
clique $A(d,0)$. Thus, the diameter equals to 3.

{\em Further steps}. Clearly, at any step $t\geq d+2$, the diameter
always lies between a pair of vertices that have just been created
at this step. We will call the newly created vertices ``outer''
vertices. At any step $t\geq d+2$, we note that an outer vertex
cannot be connected with two or more vertices that were created
during the same step $0<t'\leq t-1$. Moreover, by construction no
two vertices that were created during a given step are neighbors,
thus they cannot be part of the same $(d+1)$-clique. Thus, for any
step $t\geq d+2$, some outer vertices are connected with vertices
that appeared at pairwise different steps. Thus, there exists an
outer vertex $v_t$ created at step $t$, which is connected to
vertices $v_i's$, $1\leq i\leq t-1$, all of which are pairwise
distinct. We conclude that $v_t$ is necessarily connected to a
vertex that was created at a step $t_0\le t-d-1$. If we repeat this
argument, then we obtain an upper bound on the distance from $v_t$
to the initial clique $A(d,0)$. Let $t=\alpha (d+1)+\beta$, where
$1\leq \beta\leq d+1$. Then, we see that $v_t$ is at distance at
most $\alpha +1$ from a vertex in $A(d,0)$. Hence any two vertices
$v_t$ and $w_t$ in $A(d,t)$ lie at distance at most $2(\alpha
+1)+1$; however, depending on $\beta$, this distance can be reduced
by 1, since when $\beta\leq \lceil\frac{d}{2}\rceil$, we know that
two vertices created at step $\beta$ share at least a neighbor in
$A(d,0)$. Thus, when $1\leq p\leq \lceil\frac{d}{2}\rceil$,
$diam(A(d,t))\leq 2(\alpha +1)$, while when $\lceil\frac{d}{2}\rceil
+1\leq p\leq d+1$, $diam(A(d,t))\leq 2(\alpha +1)+1$.
One can see that these distance bounds can be reached by pairs of
outer vertices created at step $t$. More precisely, those two
vertices $v_t$ and $w_t$ share the property that they are connected
to $d$ vertices that appeared respectively at steps $t-1,t-2,\ldots
t-d-1$.

Based on the above arguments, one can easily see that for $t>d+2$,
the diameter increases by 2 every $d+1$ steps. More precisely, we
have the following result, for any $d\geq 1$ and $t\geq 1$ (when
$t=0$, the diameter is clearly equal to 1):
$$diam(A(d,t))=2(\lfloor\frac{t-1}{d+1}\rfloor +1)+f(d,t)$$
where $f(d,t)=0$ if $t-\lfloor\frac{t-1}{d+1}\rfloor (d+1)\leq
\lceil\frac{d}{2}\rceil$, and 1 otherwise.

In the limit of large $t$, $diam(A(d,t))\sim \frac{2t}{d+1}$, while
$N_t\sim [m(d+1)]^{t}$, thus the diameter is small and scales
logarithmically with the network order.

\section{Synchronization on some limiting cases for evolutionary Apollonian networks}

The ultimate goal of the study of network structure is to study and
understand the workings of systems built upon those
networks\cite{Ba02,DoMe03,SaVe04,Ne03}. Recently, some researchers
have focused on the analysis of functional or dynamical aspects of
processes occurring on networks. One particular issue attracting
much attention is the synchronizability of oscillator coupling
networks \cite{St03}. Synchronization is observed in diverse natural
and man-made systems and is directly related to many specific
problems in a variety of different disciplines. It has found
practical applications in many fields including communications,
optics, neural networks and geophysics
\cite{PeCa90,WiRa90,HaSo92,CuOp93,Vi99,Ot00}. After studying the
relevant characteristics of network structure, which is described in
the last sections, we will study the synchronization behavior on the
networks.

We follow the general framework proposed in \cite{BaPe02,PeBa98},
where a criterion based on spectral techniques was established to
determine the stability of synchronized states on networks. Consider
a network of $N$ identical dynamical systems with linearly and
symmetric coupling between oscillators. The set of equations of
motion for the system are
\begin{equation}
\dot{\textbf{x}}_i=\textbf{F}(\textbf{x}_i)+\sigma\sum_{j=1}^NG_{ij}\textbf{H}(\textbf{x}_j),
\end{equation}
where $\dot{\textbf{x}}_i=\textbf{F}(\textbf{x}_i)$ governs the
dynamics of each individual node, $\textbf{H}(\textbf{x}_j)$ is the
output function and $\sigma$ the coupling strength, and $G_{ij}$ is
the Laplacian matrix, defined by $G_{ii}=k_i$ if the degree of node
$i$ is $k_i$, $G_{ij}=-1$ if nodes $i$ and $j$ are connected, and
$G_{ij}=0$ otherwise.

Since matrix $G$ is positive semidefinite and each rows of it has
zero sum, all eigenvalues of $G$ are real and non-negative and the
smallest one is always equal to zero. We order the eigenvalues as
$0=\lambda_1\leq\lambda_2\leq\cdots\leq\lambda_{N}$. Then one can
use the ratio of the maximum eigenvalue $\lambda_{N}$ to the
smallest nonzero one $\lambda_2$ to measure the synchronizability of
the network \cite{BaPe02,PeBa98}. If the eigenratio
$R=\lambda_{N}/\lambda_2$ satisfies $R<\alpha_2/\alpha_1$, we say
the network is synchronizable. Here the eigenratio $R$ depends on
the the network topology, while $\alpha_2/\alpha_1$ depends
exclusively on the dynamics of individual oscillator and the output
function. Ratio $R=\lambda_{N}/\lambda_2$ represents the
synchronizability of the network: the larger the ratio, the more
difficult it is to synchronize the oscillators, and vice versa.

\begin{figure}
\begin{center}
\includegraphics[width=0.45\textwidth]{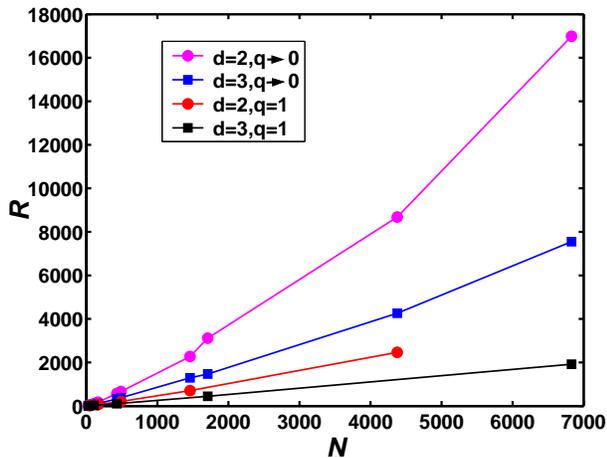} \
\end{center}
\caption[kurzform]{\label{Fig5} (Color online) The eigenratio $R$ as
a function of network order $N$. All quantities for random networks
are averaged over 50 realizations.}
\end{figure}

After reducing the issue of synchronizability to finding eigenvalues
of the Laplacian matrix $G$, we now investigate the synchronization
of our networks. Here we only study two limiting cases:
$q=1$~\cite{AnHeAnSi05,DoMa05,ZhCoFeRo05} and $q\rightarrow0$ (but
$q \neq0$)~\cite{ZhYaWa05,ZhRoCo05}. The eigenratio $R$ of different
networks is obtained numerically for different $d$ and $q$, as
exhibited in Fig.~(\ref{Fig5}). One can see that for the same $d$
and $N$, $R$ of the deterministic networks is smaller than that of
their corresponding random versions, which implies that the
synchronizability of the former is better. Even in the case of
different $d$, the deterministic networks ($d=2$) are easier to
synchronize than those random networks ($d=3$) with the same order
but different average node degree.

Why coupling systems on the two class of networks exhibit very
different synchronizability? Previously reported results have
indicated that underlying network structures play significant roles
in the synchronizability of coupled oscillators. However, the key
structural feature that determines the collective synchronization
behavior remains unclear. Many works have discussed this issue. Some
authors believe that shorter APL tends to enhance synchronization
\cite{BaPe02,GaHu00,ZhZhWa05}. In contrast, Nishikawa et. al.
reported that synchronizability is suppressed as the degree
distribution becomes more heterogeneous, even for shorter APL
\cite{NiMoLaHo03}. Also, some investigations showed that $R$
decreases whenever the betweenness heterogeneity decreases
\cite{HokiCh04}, while an opposite conclusion was claimed in
\cite{ZhZhWaYa05}. In \cite{GoYaYa05}, the authors asserted that
larger average node degree corresponds to better synchroizability.
All these may rationally explain the relations between
synchroizability and network structure in some cases, but do not
well account for the synchroizability on the evolutionary Apollonian
networks. More recently, some other authors have presented that
structure and distribution of hubs is the key in what refers to
enhance synchroizability \cite{DoHuMu05,ZhKu06}, which may be a
possible explanation for the better synchroizability of
deterministic networks when compared to random ones. But we
speculate that the synchroizability on the evolutionary Apollonian
networks is not determined  by a single structure property, but by
the combination of APL, heterogeneity of degree distribution,
betweenness centrality, modularity, mean node degree, and so on
\cite{DoHuMu05}, which need further deep research.

\section{Conclusion and discussion}

In summary, on the basis of Apollonian packings, we have proposed
and studied two kinds of evolving networks: evolutionary Apollonian
networks (EAN) and general deterministic Apollonian networks (GDAN).
According to the network construction processes we have presented
two algorithms to generate the networks, based on which we have
obtained the analytical and numerical results for degree
distribution and clustering coefficient, as well as the average path
length, which agree well with large amount of real observations. The
degree exponents can be adjusted continuously between 2 and 3, and
the clustering coefficient is very large. Moreover, we have studied
the synchronization of some limiting cases of the EAN and found that
the stochastic networks are more hardly synchronized than their
deterministic counterparts.

Because of their three important properties: power-law degree
distribution, small intervertex separation and large clustering
coefficient, the proposed networks possess good structural features
in accordance with a variety of real-life networks. For the special
case of $d=2$, the networks are maximal planar graphs. This may be
helpful for designing printed circuits. Moreover, our networks
consist of complete graphs, which has been observed in variety of
the real-world networks, such as movie actor collaboration networks,
scientific collaboration networks and networks of company
directors~\cite{Ba02,DoMe03,SaVe04,Ne03}.
\smallskip

This research was supported in part by the National Natural Science
Foundation of China (NNSFC) under Grant Nos. 60373019 and 90612007.
Lili Rong gratefully acknowledges partial support from NNSFC Grant
Nos. 70431001 and 70571011. The authors thank Dr. Wen-Xu Wang for
his assistance in preparing the manuscript, as well as the anonymous
referees for their valuable comments and suggestions.

\end{document}